\documentstyle[12pt]{article}
\newcommand {\beq}{\begin{equation}}
\newcommand {\eeq}{\end{equation}}
\newcommand {\beqa}{\begin{eqnarray}}
\newcommand {\eeqa}{\end{eqnarray}}
\newcommand {\beqan}{\begin{eqnarray*}}
\newcommand {\eeqan}{\end{eqnarray*}}
\newcommand {\n}{\nonumber \\}

\newcommand {\Romannumeral}[1]{\uppercase\expandafter{\romannumeral#1}}

\newcommand {\mrc} {\mbox{\scriptsize RC}}

\newcommand {\dd}{\mbox{d}}

\newcommand {\del}{\partial}

\begin{document}
\setlength{\oddsidemargin}{0cm}
\setlength{\baselineskip}{7mm}  %7mm

\begin{titlepage}

    \begin{normalsize}
     \begin{flushright}
UT-705 \\
Jan 1996
     \end{flushright}
    \end{normalsize}

    \begin{Large}
       \vspace{2cm}
       \begin{center}
         {\Large The Continuum Limit of One-Dimensional Quantum Regge Calculus
with Massive Bosons} \\       \end{center}
    \end{Large}

  \vspace{1cm}

\begin{center}
Takayuki Nakajima \footnote {E-mail address :
nakajima@danjuro.phys.s.u-tokyo.ac.jp,nakajima@theory.kek.jp}\\
      \vspace{1cm}
$\ast$ {\it Department of Physics, The University of Tokyo,}\\
	{\it Bunkyo-ku, Tokyo 113, Japan}
      \vspace{2cm}
\end{center}

%\hspace{5cm}

\begin{abstract}
\noindent
The most essential problems in Regge calculus discretization are
the definitions of the partition function and
the integral measure for link--length.
In recent work, by considering the one--dimensional case, it was suggested that
we should define the partition function in a certain form.
But in that work, the model which authors used was over simplified
hence the conclusions may be unreliable.
To confirm their claim, we consider a case that is more realistic.
\end{abstract}

\end{titlepage}

Recently the study of quantum gravity has made great steps.
Some of these are due to the lattice regularization scheme and
Monte-Carlo simulations. Because quantum gravity is not renormalizable
in four dimensions, we cannot treat it perturbatively
{}.
To use a non--perturbative approach we must regularize the theory.
There are two different types of lattice regularization of
quantum gravity; one is random triangulation (or dynamical triangulation)
and the other is Regge calculus. In the path integral formalism,
the quantization of geometry is performed by integrating over the
metric field. In random triangulation it is represented by
the fluctuation of the lattice structure. Random triangulation is
exactly solved in two dimensions by using the matrix model \cite{mat-mod}, and
its continuum limit is shown to reproduce Liouville theory \cite{lio-th}.
Reparametrization invariance is naturally recovered in random triangulation,
though conformal invariance is not guaranteed.
But from the viewpoint of Monte--Carlo simulation, random triangulation
is rather difficult for computers. Because of the lattice
structure changes dynamically, it is very difficult to write a vectorized
code. In four dimensions, which many people have studied for three years, it is
suggested that there exists a second order phase transition \cite{4DQG},
but it is not known whether there is a sensible
continuum limit at the critical point. We need to repeat the calculation
on a larger lattice in order to establish this.

In this context, a study of Regge calculus is worthwhile.
In Regge calculus, the lattice structure is fixed, and
metric integration is represented by integration over link--lengths
\cite{Regge}.
Because this system is nothing but an ordinary statistical system,
we can easily enlarge the system.
But unfortunately, Regge calculus has many problems.
Firstly, there is the  uncertainty of general coordinate invariance in
Regge calculus. Along with this, we have no guiding principle for choosing
the measure for integrating over link--length.
Recently, several groups have tried to test the Regge calculus
in two dimensions \cite{GH,BV,HJ,NO}, but these problems are not considered.
In \cite{BV}, using a scale--invariant measure,
they pointed out that Regge calculus does not reproduce
the value of the string susceptibility (which is a reaction of the partition
function to the variation of the volume of the universe) obtained with
continuum theory.
However, in \cite{NO}, from the measurement of the loop--length distribution,
which represents the nature of the surface,
they suggested that Regge calculus is not so discouraging
because the loop--length distribution for the ``baby loops'' has been
reproduced.
Moreover, they claimed that the definition of
the partition function in Regge calculus is subtle,
and the definition which they used might be wrong.

In this letter, considering one dimensional gravity coupled with
{\it massive} bosons, we show that Regge calculus reproduces the proper
continuum results in one dimension.
Without mass terms, it was  already shown that
it reproduces the proper continuum results in \cite{nishi}.
But it was a over simplified model.
The non--zero mass of the bosonic fields is important for two reasons.
One reason is that IR divergences are avoided.
The other reason is rather technical.
In Regge calculus, without mass terms, we can integrate link--length
( i.e. integrate over the metric field) separately, so it is not
certain that gravitational effects are present in the theory.

We consider a loop,  parametrized with
the parameter $\tau$, which runs $1 \geq \tau \geq 0$.
We put $D$ matter fields $X^\mu(\tau)$ ($\mu = 1,2,\cdots,D$)
and the metric $h(\tau)$ on the loop.
We take the action
\begin{equation}
S = \int \dd\tau \sqrt{h(\tau)} \left(
  \frac{1}{2} h^{-1}(\tau)
    \left( \frac{\del X^\mu(\tau)}{\del \tau} \right)^2
 +\frac{1}{2}M^2 X^2(\tau) \right).
\label{action}
\end{equation}
Under this action we have the partition function
\begin{eqnarray}
Z[L] &=& \int {\cal D}h {\cal D}X e^{-S} \delta ( \int d\tau \sqrt{h} -L ) \n
&=& \mbox{const.} \frac{1}{M^D L^D}
  \left( \frac{ML}{2}{\sinh (\frac{ML}{2})} \right)^D ,
%&\sim& \frac{1}{M^D L^D} \frac{1}{1+\frac{1}{24}M^2 L^2}
\end{eqnarray}
and the Green's function (or 2-point function)
\begin{eqnarray}
G^{\mu \nu} (L';L)
  &=& \left<  \left(\int^1_0 \dd\tau_1 \sqrt{h(\tau_1)} X^\mu (\tau_1) \right)
    \left( \int^1_0 \dd\tau_2 \sqrt{h(\tau_2)} X^\nu (\tau_2) \right) \right.
\n
&& \;\;\;\; \times \left.
    \delta\left(\int_0^1 \dd \tau \sqrt{h(\tau)} - L \right)
    \delta\left(\int_{\tau_1}^{\tau_2} \dd \tau \sqrt{h(\tau)} - L' \right)
    \right> \n
  &=& \eta^{\mu \nu} \frac{L}{2M} \frac{e^{-ML'}+e^{-M(L-L')}}{1-e^{-ML}} .
%  &\sim& \frac{1}{M^2} \left( 1 + M^2 \left(
%  \frac{1}{4}\left( {L'}^2 + (L-L')^2 \right) -\frac{1}{6} L^2 \right)
% \right).
\label{cont-ful-green}
\end{eqnarray}

Next, we calculate the corresponding quantity in Regge calculus.
Discretizing into $n$ pieces, the metric is re-expressed by
the link length $l_i$ ($i=1,2,\cdots,n$) between the $(i-1)$th site
and $i$th. The matter field on the $i$th site is $X_i^\mu$. We take
$l_{k+n}=l_k$ and $X_{k+n}^\mu = X_k^\mu$ for the periodic condition.
Then the action Eq.(\ref{action}) becomes
\begin{equation}
S = \sum_{i=1}^n
  \frac{\left( X_i^\mu - X_{i-1}^\mu \right)^2}{2 l_i} +
  \frac{1}{2} M^2 X_i^2 \frac{l_i+l_{i+1}}{2}.
\end{equation}
Partition functions are defined by the integral
\begin{eqnarray}
Z_0[L] &=&
  \int [\dd l] [\dd X] e^{-S} \delta\left(\sum_{i=1}^{n} l_i - L \right), \n
Z_2^{\mu \nu}[L';L] &=&
     \int [\dd l] [\dd X] e^{-S}
  \left( \sum_{j=1}^{n} \frac{l_{j}+l_{j+1}}{2} X_j^\mu \right)
  \left( \sum_{k=1}^{n} \frac{l_{k}+l_{k+1}}{2} X_k^\nu \right)
    \delta\left(\sum_{i=1}^{n} l_i - L \right)
    \delta\left(\sum_{i=j}^{k} l_i - L' \right) \n
&=& L \int [\dd l] [\dd X] e^{-S}
    \sum_{n'=1}^{n-1} \frac{l_{n'}+l_{n'+1}}{2} X_0^\mu X_{n'}^\nu
    \delta\left(\sum_{i=1}^{n} l_i - L \right)
    \delta\left(\sum_{i=1}^{n'} l_i - L' \right) ,
\label{re-part}
\end{eqnarray}
where $[\dd l]=\prod_{i=1}^{n} \dd l_i \rho_n(l_i),$ and
$[\dd X]=\prod_{i=1}^{n} \dd^D X_i^\mu$.
We should take the measure of the link--length
integration, $\rho_n(l_i),$  as
\begin{eqnarray}
\rho_n(l_i)(2\pi l_i)^{D/2} &=& \frac{\lambda^N e^{- \lambda l_i}}{\Gamma(N)}
l_i^{N-1},  \label{measure} \\
 \frac{1}{n} << &N& << n, \n
 Nn/\lambda = &\bar{L}& = \mbox{const.}
\label{scaling}
\end{eqnarray}
The definition of the continuum limit of the partition function should be
\begin{equation}
Z^{(c)}[\bar{L}]=
  \lim_{n \rightarrow \infty} \int^\infty_0 \dd L \;Z^{(\mrc)}[L].
\label{cont-lim-part}
\end{equation}
Likewise, the Green's function is defined as
\begin{equation}
G^{\mu \nu}(L';L) = \frac{Z_2^{\mu \nu}[L';L]}{Z_0[L]}.
\end{equation}

To calculate the partition functions of Eq.(\ref{re-part}),
we should begin by evaluating the matter integral
\begin{equation}
\int \prod_{i=1}^{n} \dd^D X_i^\mu  f(X) e^{-S},
\label{matter-int}
\end{equation}
where $f(X)$ is $1$ or $X_0^\mu X_{n'}^{\nu}$.
Unfortunately this integral is very complicated, and
we cannot calculate it in exactly.
So, we assume that $M$ is very small and expand in $M$.
After we integrate out $X_1$ to $X_{k-1}$ ($k \leq n'$),
Eq.(\ref{matter-int}) becomes
\begin{equation}
\left( \frac{\prod_{i=1}^{k} 2 \pi l_i}{2 \pi \tilde{L}_i} \right)^{D/2}
\int \prod_{i=k}^{n} \dd^D X_i^\mu f(X) \exp \left(
  - \left( \frac{1}{4}M^2 a_k X_0^2 + \frac{(X_k^\mu -X_0^\mu)^2}{2
\tilde{L}_k}
    + \frac{1}{4}M^2 b_k X_k^2 +\cdots \right) \right).
\end{equation}
We can find recursion relations for $a_i$, $b_i$ and $\tilde{L}_i$ as
\begin{eqnarray}
\tilde{L}_i &=& \tilde{L}_{i-1} + l_i + \frac{1}{2} M^2 \tilde{L}_{i-1} l_i
b_{i-1}, \n
a_i &=& a_{i-1} + \frac{l_i}{\tilde{L}_i} b_{i-1}, \n
b_i &=& l_i + l_{i+1} + \frac{\tilde{L}_{i-1}}{\tilde{L}_i} b_{i-1}.
\end{eqnarray}
To solve these equations, we expand $\tilde{L}_i$, $a_i$ and $b_i$ in M
\begin{eqnarray}
\tilde{L}_i &=& \tilde{L}^{(0)}_i + M^2 \tilde{L}^{(1)}_i + \cdots, \n
a_i &=& a^{(0)}_i + M^2 a^{(1)}_i + \cdots, \n
b_i &=& b^{(0)}_i + M^2 b^{(1)}_i + \cdots,
\end{eqnarray}
and we solve order by order.
The recursion relations for the leading terms,
$\tilde{L}^{(0)}_i,$ $a^{(0)}_i$ and $b^{(0)}_i,$ are easily solved as
\begin{eqnarray}
\tilde{L}^{(0)}_i &=& \sum^i_{j=1} l_j, \n
a^{(0)}_i &=& l_0 + \sum^i_{j=1} l_j, \n
b^{(0)}_i &=& \sum^i_{j=1} l_j + l_{i+1}.
\label{eq14}
\end{eqnarray}
By using Eq.(\ref{eq14}), the recursion relations for the second--order are
expressed as (for brevity we use $L_i=\sum_{j=1}^i l_j$.)
\begin{eqnarray}
\tilde{L}^{(1)}_i &=& \tilde{L}^{(1)}_{i-1} + \frac{1}{2} L_{i-1} L_i l_i, \n
a^{(1)}_i &=& a^{(1)}_{i-1}
  + \frac{l_i}{L_i} (b^{(1)}_{i-1} - \tilde{L}^{(1)}_i), \n
b^{(1)}_i &=& \frac{L_{i-1}}{L_i} b^{(1)}_{i-1} + \tilde{L}^{(1)}_{i-1}
  + \frac{L_{i-1}}{L_i} \tilde{L}^{(1)}_i.
\end{eqnarray}
Then we have
\begin{eqnarray}
\tilde{L}^{(1)}_i &=& \sum_{j_1 \neq j_2}l_{j_1}^2 l_{j_2}
    + 2 \sum_{j_1>j_2>j_3} l_{j_1} l_{j_2} l_{j_3} \n
  &\sim& \frac{1}{6} L_i^3, \n
a^{(1)}_i &=& - \frac{1}{2 L_i} \left(
  \sum_{j_1>j_2} l_{j_1}^3 l_{j_2} + \sum_{j_1>j_2} l_{j_1}^2 l_{j_2}^2
  + \sum_{j_1 \neq j_2,j_3} \sum_{j_2>j_3} 2 l_{j_1}^2 l_{j_2} l_{j_3}
  + \sum_{j_1>j_2>j_3>j_4} 4 l_{j_1} l_{j_2} l_{j_3} l_{j_4}
    \right) \n
  &\sim& - \frac{1}{12} L_i^3, \n
b^{(1)}_i &=& \frac{1}{2 L_i} \left(
  \sum_{j_1>j_2} l_{j_1} l_{j_2}^3 + \sum_{j_1>j_2} l_{j_1}^2 l_{j_2}^2
  + \sum_{j_1 \neq j_2,j_3} \sum_{j_2>j_3} 2 l_{j_1}^2 l_{j_2} l_{j_3}
  + \sum_{j_1>j_2>j_3>j_4} 4 l_{j_1} l_{j_2} l_{j_3} l_{j_4}
    \right) \n
  &\sim& - \frac{1}{12} L_i^3 .
\end{eqnarray}
To evaluate the errors in these approximations,
we consider the number of terms in each expression.
As a result, the difference between the r.h.s. and the l.h.s. is
${\cal O}(1/i^2)$. In the continuum limit, the case $i>>1$ becomes dominant.
Hence the approximation is good.

Now we can integrate out $X_1$ to $X_{n'-1}$ and $X_{n'+1}$ to $X_{n-1}$
in Eq.(\ref{matter-int}).
\begin{eqnarray}
\int \prod_{i=1}^{n} \dd^D X_i^\mu  f(X) e^{-S}
  &=& \left(\frac{\prod_{j=1}^n 2 \pi l_j}{2 \pi \tilde{L}_1 2 \pi \tilde{L}_2}
    \right)^{D/2}
  \int \dd^D X_0^\mu \dd^D X_{n'}^\mu f(X) \n
&& \;  \exp \left[ - \left(
  \frac{1}{4}M^2 \alpha (X_0^2 + X_{n'}^2)+
  (\frac{1}{2 \tilde{L}_1} + \frac{1}{2 \tilde{L}_2}) (X_0-X_{n'})^2
  \right) \right],
\end{eqnarray}
where
\begin{eqnarray}
\alpha &=& L - \frac{1}{12} M^2 \left( {L_1}^3 + {L_2}^3 \right), \n
\tilde{L}_1 &=& L_1 + \frac{1}{6} M^2 {L_1}^3, \n
\tilde{L}_2 &=& L_2 + \frac{1}{6} M^2 {L_2}^3, \n
L&=&L_1+L_2 , \;\; L_1=\sum_{i=1}^{n'} l_i , \;\; L_2=\sum_{i=n'+1}^n l_i.
\label{tris}
\end{eqnarray}
Through order $M^2$ they are
\begin{eqnarray}
\int \prod_{i=1}^{n} \dd^D X_i^\mu e^{-S}
&=& \frac{1}{M^DL^D}
  \frac{\left(\prod_{i=1}^n l_i\right)^{D/2}}
  {\left(1+\frac{1}{24}M^2 L^2\right)^{D}}, \n
\int \prod_{i=1}^{n} \dd^D X_i^\mu  X_0^\mu X_{n'}^\nu e^{-S}
&=& \frac{1}{M^{D+2}L^{D+1}}
  \frac{\left(\prod_{i=1}^n l_i\right)^{D/2}}
  {\left(1+\frac{1}{24}M^2 L^2\right)^{D}}, \n
&& \;\;\;\; \times
    \frac{1+M^2(\frac{1}{4}({L_1}^2+{L_2}^2) -
\frac{1}{12}L^2)}{1+\frac{1}{12}M^2 L^2}
 \eta^{\mu \nu}.
\end{eqnarray}

After all, the partition functions in Regge calculus are given by
\begin{eqnarray}
Z_0[L] &=& \frac{1}{M^DL^D}
  \frac{1}{\left(1+\frac{1}{24}M^2 L^2\right)^{D}}
\int [\dd l]\left(\prod_{i=1}^n l_i\right)^{D/2}
    \delta\left(\sum_{i=1}^{n} l_i - L \right),
\label{z0-bunri} \\
Z_2^{\mu \nu}[L';L] &=& \frac{1}{M^{D+2}L^D}
  \frac{1}
  {\left(1+\frac{1}{24}M^2 L^2\right)^{D}}
    \frac{1+ M^2(\frac{1}{4}({L'}^2+({L-L'})^2)
-\frac{1}{12}L^2)}{1+\frac{1}{12}M^2 L^2}
 \eta^{\mu \nu} \n
&& \;\;\;\; \times
\int [\dd l] \sum_{n'=1}^{n} \left(\prod_{i=1}^n l_i\right)^{D/2}
    \delta\left(\sum_{i=1}^{n'} l_i - L' \right)
    \delta\left(\sum_{i=n'}^n l_i - \left(L-L'\right) \right).
\label{z2-bunri}
\end{eqnarray}
(Here because of the $\delta$-functions,
we have replaced $L_1$ and $L_2$ in Eq.(\ref{tris}) with $L'$ and $L-L'$.)

Using the measure defined by Eq.(\ref{measure}),
the integral in Eq.(\ref{z0-bunri}) becomes
\begin{equation}
\frac{\lambda^{Nn} e^{-\lambda L}}{\Gamma(Nn)} L^{Nn-1},
\label{denominator}
\end{equation}
and the integral in Eq.(\ref{z2-bunri}) yields
\begin{equation}
\frac{N}{2} \frac{\lambda^{Nn} e^{-\lambda L}}{L'(L-L')}
  \sum_{n'=1}^{n-1}
\frac{{L'}^{Nn'}(L-L')^{N(n-n')}}{\Gamma(Nn')\Gamma(N(n-n'))}
  \left\{ \frac{L'}{Nn'}+\frac{L-L'}{N(n-n')} \right\}.
\label{numerator}
\end{equation}
Because $Nn>>1$ we may use Stirling's formula, and
Eq.(\ref{numerator}) becomes
\begin{equation}
\frac{N^2n}{4 \pi} \frac{\lambda^{Nn} e^{-\lambda L}}{L'(L-L')}
  e^{-Nn \ln Nn + Nn} \sum_{n'=1}^{n-1}
  \left( L'\sqrt{\frac{1-x}{x}}+(L-L')\sqrt{\frac{x}{1-x}} \right) e^{-Nn
f(x)},
\end{equation}
where $x=n'/n,$ and
\begin{equation}
f(x) = x \ln \frac{x}{L'} + (1-x) \ln \frac{1-x}{L-L'}.
\end{equation}
When $\frac{1}{\sqrt{Nn}} >> \frac{1}{n}$, the summation over $n'$ can be
evaluated by integration,
\begin{equation}
\sum_{n'=1}^{n-1} \rightarrow n \int_0^1 \dd x.
\end{equation}
So, Eq.(\ref{numerator}) is evaluated by
\begin{eqnarray}
&& \frac{N^2n^2}{4\pi}\frac{\lambda^{Nn}e^{-\lambda L}}{L'(L-L')}
   e^{-Nn\ln Nn+Nn} \n
&& \;\;\;\; \times
   \left( L'\sqrt{\frac{1-L'/L}{L'/L}}+(L-L')\sqrt{\frac{L'/L}{1-L'/L}} \right)
  L^{Nn} \sqrt{\frac{2\pi}{Nn}\frac{L'}{L}\left( 1-\frac{L'}{L}\right) } \n
&\sim& (Nn) \frac{\lambda^{Nn}e^{-\lambda L}}{\Gamma(Nn)} L^{Nn-1}.
\end{eqnarray}
Under the condition of Eq.(\ref{scaling}), Eq.(\ref{denominator}) tends to
$\delta(L-\bar{L})$. $\sqrt{Nn}$ is regarded as a
wave--function--renormalization factor.
Therefore, the continuum limit of the partition functions is given by
\begin{eqnarray}
Z^{(c)}_0[\bar{L}] &=& \int_0^\infty \dd L \; Z_0[L], \n
&=&  \frac{1}{M^D\bar{L}^D}
  \frac{1}{\left(1+\frac{1}{24}M^2 \bar{L}^2\right)^{D}}, \\
{Z^{\mu \nu}_2}^{(c)}[L';\bar{L}] &=& \int_0^\infty \dd L \; Z^{\mu
\nu}_2[L';L] \n
&=& \frac{1}{M^{D+2}\bar{L}^D}
  \frac{1}{\left(1+\frac{1}{24}M^2 \bar{L}^2\right)^{D}} \n
&& \;\;\;\;
    \frac{1+M^2(\frac{1}{4}({L'}^2+({\bar{L}-L'})^2)-\frac{1}{12}\bar{L}^2)}
      {1+\frac{1}{12}M^2 \bar{L}^2} \eta^{\mu \nu}.
\end{eqnarray}
And thus the Green's function is
\begin{eqnarray}
G^{\mu \nu}(L';\bar{L})&=&\frac{{Z^{\mu \nu}_2}^{(c)}[L';L]}{{Z_0}^{(c)}[L]} \n
&=& \frac{1}{M^2}\left(
1+M^2\left(\frac{1}{4}({L'}^2+(\bar{L}-L')^2)-\frac{1}{6}\bar{L}^2\right)\right)
\eta^{\mu \nu}.
\end{eqnarray}

So, we have properly reproduced both the partition function and the
Green's function.

In this paper, we considered massive bosons coupled to gravity in one
dimension.
After integrating out the matter field,
link--length variables are related in a complex form, at least at the
second order in the mass expansion in the continuum limit.
By using a specific integral measure and definition of the partition function,
we could reproduce the continuum Green's function.
This integral measure and the definition of the partition function are
also used in the massless case.
These results suggest that the Regge calculus is successful, at least in
one dimension. We think that there is no reason to jump to the conclusion
that Regge calculus fails, hence it is worthy of further study.

\begin{center}
\begin{large}
Acknowledgements
\end{large}
\end{center}

I would like to thank H.Kawai and J.Nishimura for helpful discussion.
I'm also grateful to R. Szalapski for carefully reading the manuscript.

\end{document}